\documentclass[prl,twocolumn,amsmath,amssymb,superscriptaddress,showpacs]{revtex4-1}

\usepackage{bm}
\usepackage{times}
\usepackage{graphicx}
\newcommand{\ket}[1]{\lvert #1 \rangle}

\newcommand{\bracket}[1]{\langle #1 \rangle}

\newcommand{\up}{\uparrow}
\newcommand{\dn}{\downarrow}

\usepackage[colorlinks,citecolor=blue,linkcolor=blue]{hyperref}

\begin{document}

\title{Coupled spin and valley physics in monolayers of MoS$_2$ and other group-VI dichalcogenides}

\author{Di Xiao}
\email{Corresponding author: xiaod@ornl.gov}
\affiliation{Materials Science and Technology Division, Oak Ridge National Laboratory, Oak Ridge, Tennessee, 37831, USA}

\author{Gui-Bin Liu}
\affiliation{Department of Physics and Center of Theoretical and Computational Physics, The University of Hong Kong, Hong Kong, China}

\author{Wanxiang Feng}
\affiliation{Materials Science and Technology Division, Oak Ridge National Laboratory, Oak Ridge, Tennessee, 37831, USA}
\affiliation{Department of Physics and Astronomy, University of Tennessee, Knoxville, Tennessee 37996, USA}
\affiliation{Institute of Physics and Beijing National Laboratory for Condensed Matter Physics, Chinese Academy of Sciences, Beijing 100190, China}

\author{Xiaodong Xu}
\affiliation{Department of Physics, University of Washington, Seattle, Washington 98195, USA}
\affiliation{Department of Material Science and Engineering, University of Washington, Seattle, Washington 98195, USA}

\author{Wang Yao}
\email{Corresponding author: wangyao@hkucc.hku.hk}
\affiliation{Department of Physics and Center of Theoretical and Computational Physics, The University of Hong Kong, Hong Kong, China}

\begin{abstract}
We show that inversion symmetry breaking together with spin-orbit coupling leads to coupled spin and valley physics in monolayers of MoS$_2$ and other group-VI dichalcogenides, making possible controls of spin and valley in these 2D materials.  
The spin-valley coupling at the valence band edges suppresses spin and valley relaxation, 
as flip of each index alone is forbidden by the valley contrasting spin splitting.  Valley Hall and spin Hall effects coexist in both electron-doped and hole-doped systems.  Optical interband transitions have frequency-dependent polarization selection rules which allow selective photoexcitation of carriers with various combination of valley and spin indices.  Photo-induced spin Hall and valley Hall effects can generate long lived spin and valley accumulations on sample boundaries.  The physics discussed here provides a route towards the integration of valleytronics and spintronics in multi-valley materials with strong spin-orbit coupling and inversion symmetry breaking.
\end{abstract}


\maketitle

Since the celebrated discovery of graphene~\cite{novoselov2004,novoselov2005a,zhang2005}, there has been a growing interest in atomically thin two-dimensional (2D) crystals for potential applications in next-generation nano-electronic devices~\cite{novoselov2005,2Dcrystal}.  Layered transition-metal dichalcogenides represent another class of materials that can be shaped into monolayers~\cite{novoselov2005}, which display distinct physical properties from their bulk counterpart~\cite{splendiani2010,mak2010,MoS2transistor,photocarrierdynamicsMoS2}.  Recent experiments have demonstrated that MoS$_2$, a prototypical group-VI dichalcogenide, crossovers from an indirect-gap semiconductor at multilayers to a direct band-gap one at monolayer~\cite{splendiani2010,mak2010}.  The direct band-gap is in the visible frequency range, most favorable for optoelectronic applications.  Monolayer MoS$_2$ transistor was also realized, demonstrating a room-temperature mobility over 200 cm$^2$/(V$\cdot$s)~\cite{MoS2transistor}.

In monolayer MoS$_2$, the conduction and valence band edges are located at the corners ($K$ points) of the 2D hexagonal Brillouin zone~\cite{abinitio2Dcrystal1,abinitio2Dcrystal2,zhu2011}.  Similar to graphene, the two inequivalent valleys constitute a binary index for low energy carriers.
% which may be utilized in electronic applications~\cite{gunawan2006,rycerz2007,xiao2007,yao2008,zhang2011,zhu2011a}. 
Because of the large valley separation in momentum space, the valley index is expected to be robust against scattering by smooth deformations and long wavelength phonons. The use of valley index as a potential information carrier was first suggested in the studies of conventional semiconductors such as AlAs and Si~\cite{gunawan2006}. With the emergence of graphene, the concept of valleytronics based on manipulating the valley index has attracted great interests~\cite{rycerz2007,xiao2007,yao2008,zhang2011,zhu2011a}.

MoS$_2$ monolayers have two important distinctions from graphene.  First, inversion symmetry is explicitly broken in monolayer MoS$_2$, which can give rise to the valley Hall effect where carriers in different valleys flow to opposite transverse edges when an in-plane electric field is applied~\cite{xiao2007}. Inversion symmetry breaking can also lead to valley-dependent optical selection rules for inter-band transitions at $K$ points~\cite{yao2008}.  Second, MoS$_2$ has a strong spin-orbit coupling (SOC) originated from the $d$-orbitals of the heavy metal atoms~\cite{zhu2011}, and can be an interesting platform to explore spin physics and spintronics applications absent in graphene due to its vanishing SOC~\cite{min2006,yao2007}.

In this Letter, we show that inversion symmetry breaking together with strong SOC lead to coupled spin and valley physics in monolayer MoS$_2$ and other group-VI dichalcogenides, making possible spin and valley control in these 2D materials.  
We find the conduction and valence band edges near $K$ points are well described by massive Dirac fermions with strong valley-spin coupling in the valence band, which has several important consequences.
First, the valley Hall effect is accompanied by a spin Hall effect in both electron-doped and hole-doped systems~\cite{murakami2003,sinova2004,kato2004,wunderlich2005}.
Second, spin and valley relaxation are suppressed at the valence band edges as flip of each index alone is forbidden by the valley-contrasting spin splitting ($\sim$ 0.1 - 0.5~eV) caused by inversion symmetry breaking.  Third, the valley-dependent optical selection rule also becomes spin-dependent, and carriers with various combination of valley and spin indices can be selectively excited by optical fields of different circular polarizations and frequencies.  We predict photo-induced charge Hall, spin Hall and valley Hall effects. The latter two phenomena can be used to generate long lived spin and valley accumulations on sample boundaries.  The physics discussed here provides a route towards the integration of valleytronics and spintronics in multi-valley materials with strong spin-orbit coupling and inversion symmetry breaking.

\begin{figure}
\includegraphics[width=8cm]{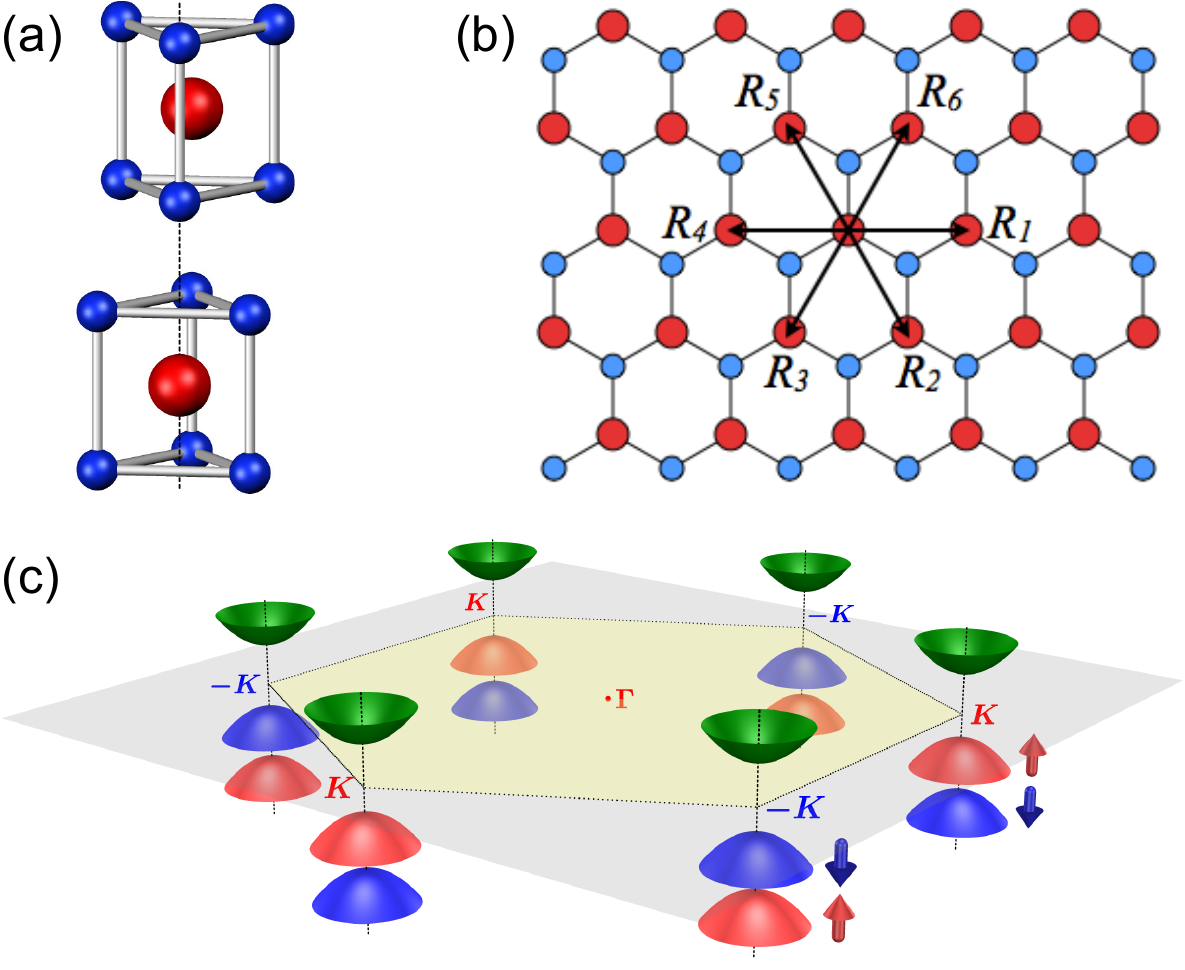}
\caption{\label{fig:xtal}(color online). (a) The unit cell of bulk $2H$-MoS$_2$, which has the inversion center located in the middle plane.  It contains two unit cells of MoS$_2$ monolayers, which lacks an inversion center. (b) Top view of the MoS$_2$ monolayer.  $\bm R_i$ are the vectors connecting nearest Mo atoms. (c) Schematic drawing of the band structure at the band edges located at the $K$ points.}
\end{figure}

The physics in monolayers is essentially the same for group-VI dichalcogenides $MX_2$ ($M$ = Mo, W, $X$ = S, Se), described below using MoS$_2$ as an example.  Structurally, MoS$_2$ can be regarded as strongly bonded 2D S-Mo-S layers that are loosely coupled to one another by Van der Waals interactions.  Within each layer, the Mo and S atoms form 2D hexagonal lattices, with the Mo atom being coordinated by the six neighboring S atoms in a trigonal prismatic geometry (Fig.~\ref{fig:xtal}a-b).  In its bulk form, MoS$_2$ has the $2H$ stacking order with the space group $D^4_{6h}$, which is inversion symmetric.  When it is thinned down to a monolayer, the crystal symmetry reduces to $D^1_{3h}$, and inversion symmetry is explicitly broken: taking the Mo atom as the inversion center, an S atom will be mapped onto an empty location.  
As a consequence, the effects we predict here are expected only in thin films with odd number of layers, since inversion symmetry is preserved in films with even number of layers.  

We start by constructing a minimal band model on the basis of general symmetry consideration.  The band structure of MoS$_2$, to a first approximation, consists of partially filled Mo $d$-bands lying between Mo-S $s$-$p$ bonding and anti-bonding bands~\cite{mattheiss1973}.  The trigonal prismatic coordination of the Mo atom splits its $d$-orbitals into three groups: $A_1(d_{z^2})$, $E(d_{xy}, d_{x^2-y^2})$ and $E'(d_{xz}, d_{yz})$.  In the monolayer limit, the reflection symmetry in the $\hat{z}$ direction permits hybridization only between $A_1$ and $E$ orbitals, which opens a band gap at the $K$ and $-K$ points~\cite{mattheiss1973}, schematically shown in Fig.~\ref{fig:xtal}c.  The group of the wave vector at the band edges ($K$) is $C_{3h}$ and the symmetry adapted basis functions are 
\begin{equation}
\ket{\phi_c} = \ket{d_{z^2}} \;, \quad
\ket{\phi_v^\tau} = \frac{1}{\sqrt{2}}(\ket{d_{x^2 - y^2}} + i\tau \ket{d_{xy}}) \;,
\end{equation}
where the subscript $c(v)$ indicates conduction (valence) band, and $\tau = \pm 1$ is the valley index.  The valence-band wave functions at the two valleys, $\ket{\phi_v^+}$ and $\ket{\phi_v^-}$, are related by time-reversal operation.  To first order in $k$, the $C_{3h}$ symmetry dictates that the two-band $k\cdot p$ Hamiltonian has the form
\begin{equation}
\hat{H}_0 = at(\tau  k_x\hat{\sigma}_x + k_y\hat{\sigma}_y) + \frac{\Delta}{2} \hat{\sigma}_z \;,
\end{equation}
where $\hat{\sigma}$ denotes the Pauli matrices for the two basis functions, $a$ is the lattice constant, $t$ the effective hopping integral, and $\Delta$ the energy gap.  These parameters are obtained by fitting to first-principles band structure calculations and are listed in Table.~\ref{fitting} for the four group-VI dichalcogenides~\footnote{We have also included second-order terms, but found they are negligible compared with the first-order terms.}. We note that the same effective Hamiltonian also describes monolayer graphene with staggered sublattice potential~\cite{xiao2007,yao2008}.  This is not surprising, as both systems have the same symmetry properties.  What distinguishes MoS$_2$ from graphene is the strong SOC originated from the metal $d$-orbitals.  The conduction band-edge state is made of $d_{z^2}$ orbitals and remains spin-degenerate at $K$ points, whereas the valence band-edge state splits.  Approximating the SOC by the intra-atomic contribution $\bm L \cdot \bm S$, we find the total Hamiltonian given by
\begin{equation} \label{ham}
\hat{H} = at(\tau  k_x\hat{\sigma}_x + k_y\hat{\sigma}_y) + \frac{\Delta}{2}\hat{\sigma}_z - \lambda \tau \frac{\hat{\sigma}_z -1}{2} \hat{s}_z \;,
\end{equation}
where $2\lambda$ is the spin-splitting at the valence band top caused by the SOC and $\hat{s}_z$ is the Pauli matrix for spin.  The spin-up ($\up$) and spin-down ($\dn$) components are completely decoupled and $s_z$ remains a good quantum number.  We emphasize that the spin splitting does not depend on the model details; it is a general consequence of inversion symmetry breaking, similar to the Dresselhaus spin splitting in zinc-blende semiconductors~\cite{dresselhaus1955}.  Time-reversal symmetry requires that the spin splitting at different valleys must be opposite (Fig.~\ref{fig:xtal}c)~\footnote{The valence band spin splitting at $K$ points is the consequence of spin orbit coupling and inversion symmetry breaking in monolayer. It is comparable in size with the valence band splitting at $K$ points seen in the bilayer and in the bulk. But the latter is from the interlayer coupling~\cite{mattheiss1973}, and the presence of both inversion symmetry and time reversal symmetry forbids any spin splitting at $K$ points.}.

\begin{table}
\caption{\label{fitting}Fitting result from first-principles band structure calculations. The monolayer is relaxed. The sizes of spin splitting $2 \lambda$ at valence band edge were previously reported in the first principle studies~\cite{zhu2011}. The unit is \AA ~for $a$, and eV for $t$, $\Delta$ and $\lambda$. $\Omega_{1}$ ($\Omega_{2}$) is the Berry curvature in unit of \AA$^2$, evaluated at $-K$ point for the spin up (down) conduction band. }
\begin{ruledtabular}
\begin{tabular}{ccccccccc}
         & $a$  & $\Delta$ & $t$ & $2\lambda$ & $\Omega_{1}$ & $\Omega_{2}$ \\ \hline
MoS$_2$  & 3.193 & 1.66 & 1.10 & 0.15 & 9.88  & 8.26 \\
WS$_2$   & 3.197 & 1.79 & 1.37 & 0.43 &15.51  & 9.57 \\
MoSe$_2$ & 3.313 & 1.47 & 0.94 & 0.18 & 10.23 & 7.96 \\
WSe$_2$  & 3.310 & 1.60 & 1.19 & 0.46 & 16.81 & 9.39 \\
\end{tabular}
\end{ruledtabular}
\end{table}

\begin{figure}
\includegraphics[width=\columnwidth]{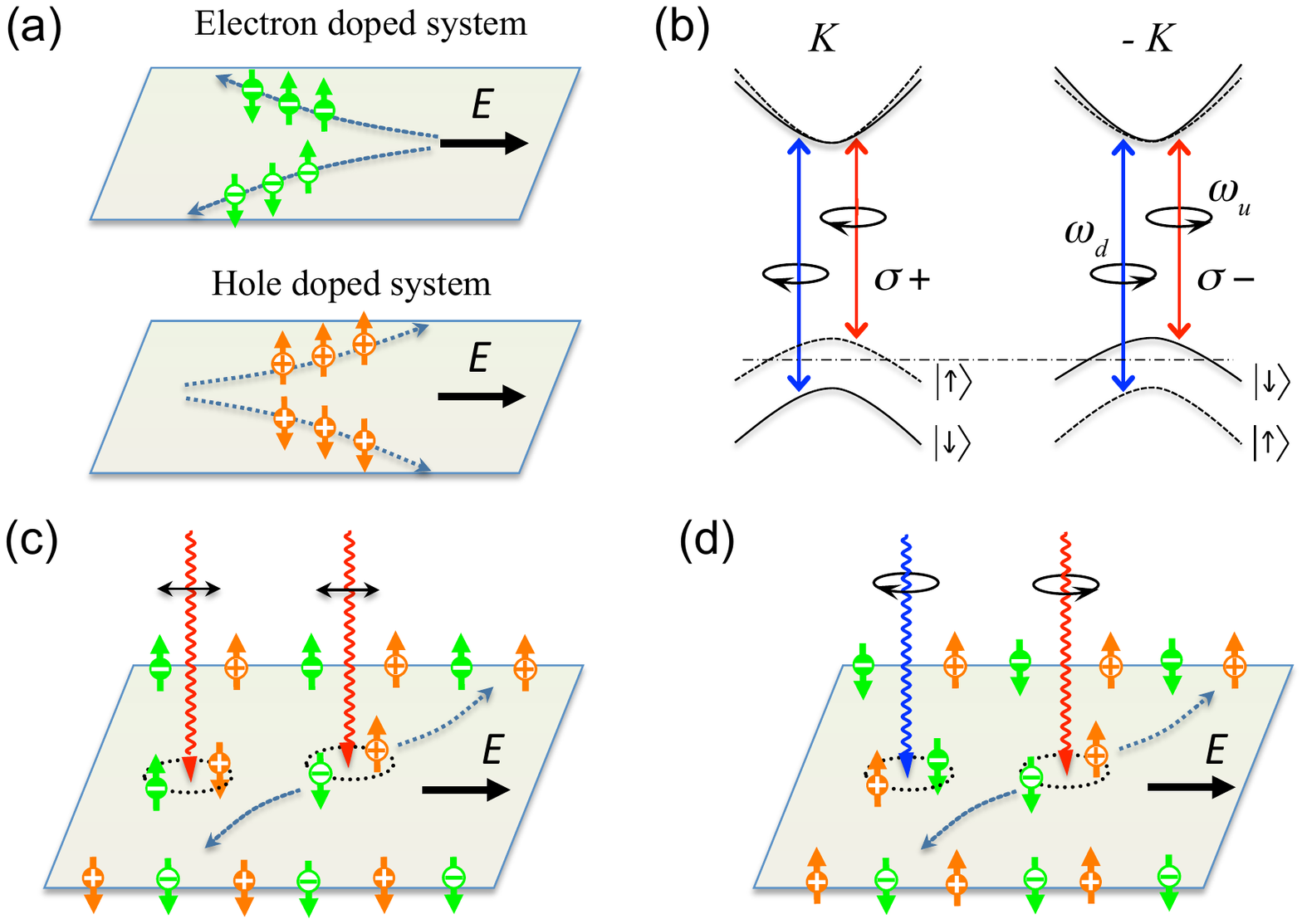}
\caption{\label{fig:optical}(color online). Coupled spin and valley physics in monolayer group-VI dichalcogenides. The electrons and holes in valley $K$ are denoted by white `$-$', `$+$' symbol in dark circles and their counterparts in valley $-K$ are denoted by inverse color. (a) Valley and spin Hall effects in electron and hole doped systems (see text). (b) Valley and spin optical transition selection rules. Solid (dashed) curves denote bands with spin down (up) quantized along the out-of-plane direction. The splitting in the conduction band is exaggerated. $\omega_u$ and $\omega_d$ are respectively the transition frequencies from the two split valence band tops to the conduction band bottom. (c) Spin and valley Hall effects of electrons and holes excited by linearly polarized optical field with frequency $\omega_u$. (d) Spin and valley Hall effects of electrons and holes excited by two-color optical fields with frequencies $\omega_u$ and $\omega_d$ and opposite circular polarizations. }
\end{figure}

The valley Hall and spin Hall effects are driven by the Berry phase associated with the Bloch electrons.  It has been well established that in the presence of an in-plane electric field, an electron will acquire an anomalous velocity proportional to the Berry curvature in the transverse direction~\cite{xiao2010}, giving rise to an intrinsic contribution to the Hall conductivity~\cite{nagaosa2010}, $\sigma^\text{int} = (e^2/\hbar) \int[d \bm k] f(\bm k) \Omega(\bm k)$, where $f(\bm k)$ is the Fermi-Dirac distribution function, and $[d\bm k]$ is a shorthand for $d\bm k/(2\pi)^2$.  The Berry curvature is defined by $\Omega_n(\bm k) \equiv \hat{\bm z} \cdot \bm\nabla_{\bm k} \times \bracket{u_n(\bm k)|i\nabla_k|u_n(\bm k)}$, where $\ket{u_n(\bm k)}$ is the periodic part of the Bloch function and $n$ is the band index.  For massive Dirac fermions described by the effective Hamiltonian in Eq.~\eqref{ham}, the Berry curvature in the conduction band is~\cite{xiao2007}:
\begin{equation}
\Omega_{c}(\bm k) = - \tau \frac{2a^2t^2\Delta'}{[\Delta'^2+4a^2t^2k^2]^{3/2}}.
\end{equation}
Note that the Berry curvatures have opposite sign in opposite valleys.
In the valence band, we have: $\Omega_{v}(\bm k) = -\Omega_{c}(\bm k)$.  In the same valley, the Berry curvature is dependent on spin through the spin-dependent band gap: $\Delta' \equiv \Delta -\tau s_z \lambda $. 
The curvature is nearly constant in the neighborhood of $K$ points since $\Delta \gg a t k$ (Table~\ref{fitting}). The valley Hall conductivity (in unit of $e/\hbar$) is then:
\begin{equation}
\sigma_{v} ^n =  2 \int[d\bm k] (f_{n, \up}(\bm k) \Omega_{n, \up}(\bm k) + f_{n, \dn}(\bm k)\Omega_{n, \dn}(\bm k)) \;, \notag
\end{equation}
and the spin Hall conductivity (in unit of $e/2$) is:
\begin{equation}
\sigma_{s}^n = 2  \int[d\bm k] (f_{n, \up} (\bm k) \Omega_{n, \up}(\bm k) - f_{n, \dn} (\bm k) \Omega_{n, \dn}(\bm k)) \;, \notag
\end{equation}
where the integration is performed over the neighborhood of one $K$ point.
For moderate hole doping with Fermi energy lying between the two split valence band tops (illustrated by the dot-dashed line in Fig.~\ref{fig:optical}b), the valley and spin Hall conductivities of holes are the same, given by
\begin{equation}
\sigma^h_{s} = \sigma^h_{v} = \frac{1}{\pi}\frac{\mu}{\Delta - \lambda}
\end{equation}
for $\mu \ll \Delta - \lambda$, where $\mu$ is the Fermi energy measured from the valence band maximum. If the system is electron doped, we must consider both conduction bands which are degenerate at $K$ points and have small spin-splitting quadratic in $k$ (see Fig.~\ref{fig:optical}b).  We find that
\begin{equation}
\sigma^e_{v} = \frac{1}{\pi} \frac{\Delta}{\Delta^2-\lambda^2}\mu \;, \quad
\sigma^e_{s} = \frac{1}{\pi} \frac{\lambda}{\Delta^2-\lambda^2}\mu \;.
\end{equation}
where $\mu$ is the Fermi energy measured from the conduction band minimum.  The spin Hall conductivity is about $\lambda/\Delta$ of the valley Hall conductivity.

The robustness of the valley and spin Hall effects is closely related to the relaxation time of the valley and spin index.  Flipping of valley index require atomic scale scatters, since the two valleys are separated by a wave vector comparable with the size of Brillouin zone. Spin flips requires the coupling with magnetic defects, since $s_z$ is a good quantum number at the conduction and valence band edges. 
In the conduction band, valley scattering could be slow in the bulk at the clean limit, but will be facilitated on the boundaries by valley mixing except with perfect zigzag edge. In the valence band, by the relatively large valley-contrasting spin splitting ($\sim$ 0.1 - 0.5~eV), valley and spin can only be simultaneously flipped to conserve energy which require atomic scale magnetic scatters. In the absence of such scatters, we expect holes have long spin and valley lifetimes both in the bulk and on the boundary. 

Next we look at optical inter-band transitions from the spin-split valence band tops to the conduction band bottoms. The coupling strength 
with optical fields of $\sigma_{\pm}$ circular polarization is given by $ \mathcal{P}_{\pm} (\bm{k}) \equiv \mathcal{P}_{x} (\bm{k}) \pm i \mathcal{P}_{y} (\bm{k})$, where $ \mathcal{P}_{\alpha} (\bm{k}) \equiv m_0 \langle
u_{c} (\bm{k}) |  \frac{1}{\hbar} \frac{\partial \hat{H}}{\partial k_{\alpha}} | u_{v} (\bm{k}) \rangle$ is the
interband matrix element of the canonical momentum operator and $m_0$ is the free electron mass~\cite{oscillatorstrength,yao2008}. For transitions near $K$ points, we find 
\begin{equation}
|\mathcal{P}_{\pm}(\bm k)|^2  = \frac{m_0^2  a^2 t^2}{\hbar^2} (1 \pm \tau \frac{\Delta'}{ \sqrt{\Delta'^2 + 4 a^2 t^2 k^2}})^2 \label{probability}.
\end{equation}
Since $\Delta' \gg a t k$, the interband transitions are then coupled exclusively with $\sigma +$ ($\sigma-$) circularly polarized optical field at the $K$ ($-K$) valley. Optical field couples only to the orbital part of the wave function and spin is conserved in the optical transitions. By the valley-contrasting spin splitting of the valence band tops, the valley optical selection rule becomes spin-dependent selection rules, as illustrated in Fig.~\ref{fig:optical}b. $\omega_d$ and $\omega_u$ denote here the two band edge excitonic transition frequencies from the spin-split valence band tops (see Fig.~\ref{fig:optical}b). Because of the spin-valley coupling and the valley optical selection rule from inversion symmetry breaking, spin and light polarization are related in the opposite ways at the two frequencies, similar to the interband transition involving heavy hole and light hole in III-V semiconductors. One may expect a sign reversal for magneto-optical effects such as Faraday rotation and Kerr rotation by spin-polarized electron when the frequency changes from $\omega_u$ to $\omega_d$. 

Selective excitation of carriers with various combination of valley and spin index becomes possible using optical fields of different circular polarizations and frequencies.
Optical field with $\sigma +$ circular polarization and frequency $\omega_u$ ($\omega_d$) can generate spin up (down) electrons and  spin down (up) holes in valley $K$, while the excitation in the $-K$ valley is simply the time reversal of the above~\footnote{We use the convention that an unoccupied spin up (down) state in the valence band is referred as a spin down (up) hole. By excitation at frequency $\omega_d$, carriers can also be excited from the upper valence band if the states are not empty, but the excitonic transition from the lower band will have dominant joint density of states~\cite{mak2010}.}. Such a spin and valley dependent selection rule can be used to generate long lived spin and valley accumulations on sample boundaries in a Hall bar geometry. 
Consider the photo-excitation of electrons and holes, which are then dissociated by an in-plane electric field, driving a longitudinal charge current (Fig.~\ref{fig:optical}c-d). The photo-excited electrons and holes will also acquire opposite transverse velocities because of the Berry curvatures in the conduction and valence band, and moved to the two opposite boundaries of the sample. This leads to Hall current of valleys, spins or charges, depending on the polarization and frequency of the optical field. In Table~\ref{hallconductivity}, we give the signs and order of magnitude estimation of the valley, spin and charge Hall currents in the clean limit. 

\begin{table}
\caption{\label{optical hall} Photo-induced spin, valley and charge Hall effects. $\sigma_{s}^{e(h)}$ and $\sigma_{v}^{e(h)}$ denote the spin and valley Hall conductivity respectively contributed by the photo-excited electrons (holes). $\sigma$ is the total charge Hall conductivity from both carriers. All conductivities are normalized by the photo-excited carrier density of electron or hole, and only intrinsic contribution is considered.}
\begin{ruledtabular}
\begin{tabular}{cccccc}
light frequency & $\sigma_{s}^e$ & $\sigma_{s}^h $ &  $\sigma_{v}^e$  &  $\sigma_{v}^h $ &  $\sigma$  \\
\& polarization &  &  &  &  &    \\
\hline
$(\omega_u, X~\textrm{or}~Y)$ & $ \frac{e}{2} \Omega_{1} $ & $\frac{e}{2} \Omega_{1} $ & $ \frac{e}{\hbar} \Omega_{1} $ & $ \frac{e}{\hbar} \Omega_{1} $ & 0\\
$(\omega_d, X~\textrm{or}~Y)$ & $-  \frac{e}{2} \Omega_{2} $ & $-  \frac{e}{2} \Omega_{2}  $ & $\frac{e}{\hbar} \Omega_{2}$ & $  \frac{e}{\hbar} \Omega_{2} $ & 0 \\
$(\omega_u, \sigma +)$ & $   \frac{e}{2} \Omega_{1}$ & $   \frac{e}{2} \Omega_{1}$ & $ \frac{e}{\hbar} \Omega_{1}$ & $  \frac{e}{\hbar} \Omega_{1}$ & $ 2 \frac{e^2}{\hbar} \Omega_{1}$ \\
$(\omega_d, \sigma -) $ & $ -\frac{e}{2} \Omega_{2}$ & $  - \frac{e}{2} \Omega_{2} $ & $ \frac{e}{\hbar} \Omega_{2}$ & $ \frac{e}{\hbar} \Omega_{2}$ & $ -2 \frac{e^2}{ \hbar} \Omega_{2}$  \label{hallconductivity}
\end{tabular}
\end{ruledtabular} 
\end{table}

Excitation with circular polarizations will generate a charge Hall current which can be detected as a voltage. The sign of the voltage is exclusively determined by the circular polarization and is independent of the frequency. Excitation with linear polarizations has more interesting consequences. For example, by excitation with linearly polarized optical field with frequency $\omega_u$, there is a spin Hall current and a valley Hall current in the absence of a charge Hall current. Spin up electrons from the $K$ valley and spin up holes from the $-K$ valley are accumulated on one boundary, while their time reversals are accumulated on the other boundary (Fig.~\ref{fig:optical}c). Thus, each boundary can remain charge neutral while carrying a net spin polarization as well as a net valley polarization. Recombination of these excess electrons and holes are forbidden by the optical transition selection rules unless assisted by processes which flip both the valley and spin index. Holes are expected to have much longer spin and valley lifetimes on the boundary. Thus electrons will get unpolarized first and recombine with the spin and valley polarized holes, accompanied by the emission of photons with opposite circular polarizations on the two boundaries. If there is strong valley mixing for electrons on the boundary, the decay of the overall spin and valley polarization is determined by the spin relaxation time of the electrons. 

Another interesting excitation scenario is by a non-degenerate optical excitation, consisted of a $\sigma+$ polarized component with frequency $\omega_u$ and a $\sigma-$ polarized component with frequency $\omega_d$. This will excite spin up electrons and spin down holes in both valleys. The spin Hall and charge Hall currents from the electrons will largely cancel with those from the holes, while the valley Hall currents from electrons and holes add constructively (see Table~\ref{hallconductivity}). The electrons and holes accumulated on the same boundary are of opposite spin and valley indice. When electrons get valley unpolarized, they can recombine with the spin and valley polarized holes, accompanied by the photon emission with polarization and frequency $(\sigma+, \omega_u)$ on one boundary and $(\sigma- ,\omega_d)$ on the other. This process may provide a direct measurement on the valley lifetime of electrons on the boundary.

In summary, we have predicted the valley dependent optical selection rules for interband transitions near $K$-points in monolayer MoS$_2$ and other group VI transition metal dichalcogenides. The spin-orbit interaction from the metal $d$-orbitals further leads to strong coupling of spin and valley degrees of freedom, which makes possible selective photoexcitation of carriers with various combination of valley and spin indices. We have also predicted the coexistence of valley Hall and spin Hall effects in $n$-doped and $p$-doped systems, and proposed photo-induced spin Hall and valley Hall effects for generating spin and valley accumulations on edges. The strong spin-valley coupling can further protect each index: with the valley dependent spin splitting of $O(0.1)$~eV at the valence band top, flip of spin and valley alone is energetically forbidden. These effects suggest the potential of integrated spintronic and valleytronic applications. In hybrid systems of these monolayers with other spintronics materials, spin index may be used as a universal information carrier across different materials, while valley index provides a unique ancillary information carrier in the monolayers with logic operations between the two enabled by the spin-valley coupling. 

We acknowledge useful discussions with D.~Mandrus, S.~Okamoto, and J.-Q. Yan.  We are grateful to W.-G. Zhu for technical support in first-principles band structure calculations.  This work was supported by the U.S. Department of Energy, Office of Basic Energy Sciences, Materials Sciences and Engineering Division (D.X.), by Research Grant Council of Hong Kong (G.B.L.~and W.Y.), and by the Laboratory Directed Research and Development Program of ORNL (W.F.).

Ê\textit{Note added.} -- Recently, experimental evidences on the optical selection rules for inter-band transitions at $K$ points are reported in monolayer MoS$_2$~\cite{Cui_valley}. We also note an independent theoretical work discussing the circular dichroism in the entire Brillouin zone~\cite{pku_valley}.

\end{document}